\documentclass[aps,prl,showpacs,twocolumn]{revtex4}
\usepackage{bbm}
\usepackage{mathrsfs}
\usepackage{epsfig,psfrag}
\usepackage{amsmath,amsfonts,amssymb}
\usepackage[usenames]{color}

\begin{document}

\preprint{draft}

\title{Casimir force driven ratchets}

\author{T. Emig}
\affiliation{Laboratoire de Physique Th\'eorique et Mod\`eles 
Statistiques, CNRS UMR 8626, Universit\'e Paris-Sud, 91405 Orsay, France}

\date{\today}

\begin{abstract}
  We explore the non-linear dynamics of two parallel periodically
  patterned metal surfaces that are coupled by the zero-point
  fluctuations of the electromagnetic field between them. The
  resulting Casimir force generates for asymmetric patterns with a
  time-periodically driven surface-to-surface distance a ratchet
  effect, allowing for directed lateral motion of the surfaces in
  sizeable parameter ranges. It is crucial to take into account
  inertia effects and hence chaotic dynamics which are described by
  Langevin dynamics. Multiple velocity reversals occur as a function
  of driving, mean surface distance, and effective damping.  These
  transport properties are shown to be stable against weak ambient noise.
\end{abstract}


\maketitle


The observation of Casimir forces in increasingly small devices on
sub-micron scales has generated great current interest in exploring
the role of these forces for the development and optimization of
micro- and nano-electromechanical systems
\cite{Buks:2001b,Chan:2001a,Chan:2001d}. These systems can serve as
on-chip fully integrated sensors and actuators with a growing number
of applications. It was pointed out that Casimir forces can make an
important contribution to the principal cause of malfunctions of these
devices in form of stiction that results in permanent adhesion of
nearby surface elements \cite{Buks:2001a}. This initiated interest in
repulsive Casimir forces by modifying material properties as
well as the geometry of the interacting components
\cite{Buks:2002a,Kenneth:2002b}. 

A complementary strategy, which will be considered in this Letter, is
to make actually good use of Casimir forces between metallic surfaces
and employ them to actuate components of small devices without
contact. We will demonstrate that this can be achieved by coupling two
periodically structured (locally) parallel surfaces by the zero-point
fluctuations of the electromagnetic field between them. The broken
translation symmetry parallel to the surfaces results in a sideways
force which has been predicted theoretically
\cite{Emig:2001a,Emig:2003a} and observed experimentally between
static surfaces \cite{Chen:2002a}. If at least one of the surfaces is
structured {\it asymmetrically} there is an additional breaking of
reflection symmetry and the surfaces can in principle be set into
relative lateral motion in the direction of broken symmetry. The
energy for this transport has to be pumped into the system by external
driving. This can be realized by setting the surfaces into relative
oscillatory motion so that their normal distance is an unbiased
periodic function of time. Since the sideways Casimir force decays
exponentially with the normal distance \cite{Emig:2003a}, the
surfaces experience an asymmetric periodic potential that varies
strongly in time.

This scenario resembles so-called ratchet systems \cite{Reimann:2002a}
that have been studied extensively during the last decade in the
context of Brownian particles \cite{Ajdari:1997a}, molecular motors
\cite{Astumian:1998a} and vortex physics in superconductors
\cite{Souza-Silva:2006a}, to name a few recent examples. Most of the
works on ratchets consider an external time-dependent driving force
acting on overdamped degrees of freedom to rectify thermal noise. For
nano-systems, however, it has been pointed out that inertia terms due
to finite mass should not be neglected and, actually, can help the
ratchets to perform more efficiently than their overdamped companions
\cite{Marchesoni:2006a}.  Finite inertia typically induce in Langevin
dynamics deterministic chaos that has been shown to be able to mimic
the role of noise and hence to generate directed transport in the
absence of external noise \cite{Mateos:2000a}. Here we use this effect
in the different context of so-called pulsating (or effectively
on-off) ratchets where the strengths of the periodic potential varies
in time \cite{Reimann:2002a}. We consider weak thermal noise only to
test for stability of the inertia induced transport --- not as the
source of driving. We note that in the absence of inertia, finite
thermal noise {\it is} necessary for on-off ratchets to generate
directed motion. 

In this Letter we demonstrate that the system described above indeed
allows for directed relative motion of the surfaces due to chaotic
dynamics caused by the lateral Casimir force.  The transport velocity
is stable across sizeable intervals of the amplitude and frequency of
surface distance oscillations and damping. The velocity scales linear
with frequency across these intervals and is almost constant below a
critical mean distance beyond which it drops sharply.  The system
exhibits multiple current reversals as function of the oscillation
amplitude, mean distance and damping. The Casimir ratchet allows
contact-less transmission of motion which is important since
traditional lubrication is not applicable in nano-devices.  This
actuation mechanism should be compared to other actuation schemes
as magnetomotive or capacitive (electrostatic) force transmission.
The Casimir effect induced actuation has the advantage of working also
for insulators and does not require any electrical contacts and/or
external fields. It can also scale down effects of parasitic
capacitance that reduces the efficiency of actuation at high
frequencies \cite{Buks:2001a}. It should be mentioned that other
applications of zero-point fluctuation induced (van der Waals)
interactions to nano-devices have been experimentally realized already
to construct ultra-low friction bearings from multiwall carbon
nanotubes \cite{Cumings:2000a}.

\begin{figure}
\includegraphics[width=.82\linewidth]{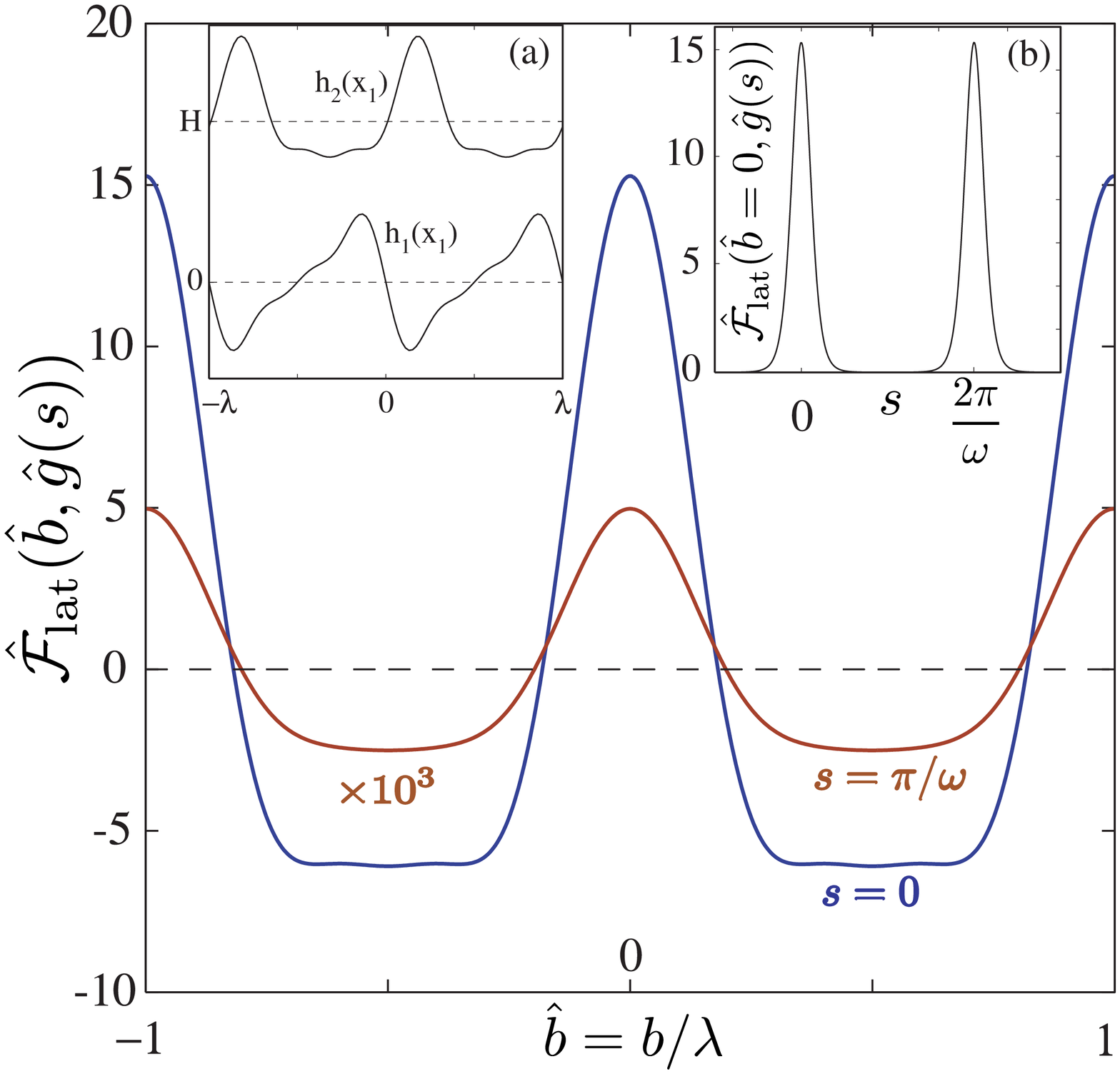}
\vspace*{-0.2cm}
\caption{\label{Fig:force} The lateral Casimir force acting between
  the two surfaces as function of the shift $\hat b$ at time $s=0$ and half
  period $s=\pi/\omega$ (drawn to a larger scale by a factor $10^3$)
  for parameters $\eta=0.65$, $H_0=0.1 \lambda$.  Insets: (a) Surface
  profiles at their equilibrium position at $\hat b=0.182$ (b)
  Periodic variation of the maximum force at $\hat b=0$ with time.}
\vspace*{-0.5cm}
\end{figure}

In the following, we consider two (on average) parallel metallic
surfaces with periodic, uni-axial corrugations (along the $x_1$-axis)
that have distance $H$, see inset (a) of Fig.~\ref{Fig:force}. To
begin with, we assume that both surfaces are at rest with a relative
lateral displacement $b$.  Then the surface profiles can be
parametrized as
\begin{subequations}
\label{corr-2}
\begin{eqnarray}
 h_1(x_1) \, &=& \,a \, \sum_{n=1}^\infty c_n
 e^{2\pi i n x_1/\lambda_1} + {\rm c.c.} \, , \\
h_2(x_1) \, &=&  \, a \,\sum_{n=1}^{\infty} d_n 
e^{2\pi i n (x_1-b)/\lambda_2} +{\rm c.c.} \, \, ,
\end{eqnarray}
\end{subequations}
where $a$ is the corrugation amplitude, $\lambda_1$, $\lambda_2$ are
the corrugation wave lengths, $c_n$, $d_n$ are Fourier
coefficients.

The Casimir energy ${\cal E}$ of this configuration is the {\it
  change} of the ground state energy of the electromagnetic field due
to the suppression of the tangential electric field at the
surfaces. The dependence of ${\cal E}$ on $H$ and $b$ causes
macroscopic forces on the surfaces.  For a varying separation $H$ this
is the normal Casimir attraction between metallic surfaces
\cite{Casimir:1948dh}, modified by the corrugations. Below we will
assume $H=H(t)$ to be a time-dependent distance that is kept at a
fixed oscillation by an additional external force from clamping to an
oscillator. In such setup, the surfaces can react freely only to the
lateral force component ${\cal F}_{\rm lat}(b,H)=-\partial {\cal
  E}/\partial b$.  This side-ways force has been computed for
sinusoidal corrugations to second order in the amplitude $a$, using a
path integral formulation \cite{Emig:2003a}.  This result is readily
extended to periodic profiles of arbitrary shapes as described by
Eq.~(\ref{corr-2}). We find that the corrugation length have to be
commensurate, $\lambda_1/\lambda_2=p/q$ with integers $p$, $q$ in
order to produce a finite lateral force per surface area. For the
purpose of this work, it is sufficient to consider the case $p=1$. The
lateral ($b$-dependent) part of the Casimir energy per surface area
can then be written as
\begin{equation}
\label{eq:lateral_energy}
{\cal E}(b)= 
\frac{2\hbar c a^2}{H^5} \sum_{n=1}^\infty \left( c_n d_{-nq} 
e^{-2\pi i n b/\lambda_1} + \text{c.c.} \right) 
J\left( n \frac{H}{\lambda_1} \right)
\end{equation}
to order $a^2$. The exact form of the function $J$ was obtained in
Ref.~\onlinecite{Emig:2003a} in terms of transcendental functions.
For the present purpose it is sufficient to use the simplified
expression
\begin{equation}
\label{eq:J-fct}
J(u)\simeq\frac{\pi^2}{120}\left(
1 + 2\pi u + \gamma  u^2 + 32 u^4 
\right)\, e^{-2\pi u}
\end{equation}
with $\gamma=12.4133$, which is exact for both asymptotically large and
small $u$ and approximates the exact results with sufficient accuracy
for all $u$. (The maximal deviation from the exact result is
$\approx\pm 0.5\%$ around $u=0.5$.) The Casimir potential of
Eq.~\eqref{eq:lateral_energy} has two interesting properties which are
useful to the construction of a ratchet. First, it decays
exponentially with $H$, and thus can be essentially switched on and
off periodically in time by oscillating $H$. Second, the potential is
not only periodic in $b$ but acquires asymmetry from the surface profiles 
at small $H\ll \lambda$ and an universal symmetric shape for $H\gg \lambda$
since the effect of higher harmonics of the surface profile is exponentially
diminished \cite{Buscher:2005a}. 

The relative surface displacement $b(t)$ can be considered as a
classical degree of freedom with inertia. Its equation of motion is
described by Langevin dynamics of the form
\begin{equation}
\label{eq:langevin}
\rho \ddot b + \gamma\rho \dot b = {\cal F}_{\rm lat}[b, H(t)] + 
\sqrt{2\gamma\rho T} \, \xi(t) \, ,
\end{equation}
where $\rho$ is the mass per surface area, $\gamma$ the friction
coefficient, $T$ the intensity (divided by surface area) of the
Gaussian noise $\xi(t)$ with zero mean and correlations $\langle
\xi(t)\xi(t')\rangle = \delta(t-t')$ so that the Einstein relation is
obeyed. This stochastic term describes ambient noise due to effects of
temperature and pressure. (Additional contributions from thermally
excited photons to the Casimir force can be neglected at surface
distances well below the thermal wavelength $\hbar c/(2 T)$.) The
system is driven by rigid oscillations of one surface so that the
distance $H(t)=H_0 g(t)$ oscillates about the mean distance $H_0$ with
$g(t)=1-\eta\cos(\Omega t)$. For simplicity, we consider now equal
corrugation lengths $\lambda_1=\lambda_2\equiv\lambda$.  We define the
following dimensionless variables: $\hat b=b/\lambda$, $s=t/\tau$ for
lateral lengths and time with the typical time scale
$\tau=(\lambda/a)\sqrt{\rho H_0^5/\hbar c}$ resulting from a balance
between inertia and Casimir force. Hence velocities will be measured
in units of $v_0=\lambda/\tau$. There are five dimensionless
parameters which can be varied independently for fixed surface
profiles: the damping $\hat \gamma=\tau\gamma$, the angular frequency
$\omega=\tau\Omega$, the driving amplitude $\eta$, the scaled mean
distance $H_0/\lambda$ and the noise intensity $\hat T=(T/\hbar
c)(H_0^5/a^2)$. The dimensionless equation of motion for $\hat b(s)$ reads
\begin{equation}
\label{eq:langevin-dimless}
\ddot{\hat b}+\hat \gamma \dot{\hat b} = \hat {\cal F}_{\rm lat}[\hat b,\hat g(s)]
+\sqrt{2\hat\gamma \hat T} \, \hat\xi(s) 
\end{equation}
with the Casimir force 
\begin{equation}
\label{eq:force-dimless}
\hat {\cal F}_{\rm lat}(\hat b, \hat g)=\frac{4\pi}{\hat g^5} 
\sum_{n=1}^\infty f_n \cos(2\pi n \hat b) J\left( n \hat g \frac{H_0}{\lambda} \right) \, ,
\end{equation}
where we have chosen surface profiles with $c_n=i\sqrt{f_n/(2n)}$,
$d_n=\sqrt{f_n/(2n)}$ with real coefficients $f_n$ in
Eq.~\eqref{corr-2}, and $\hat g(s)=1-\eta\cos(\omega s)$.


We begin to analyze Eq.~\eqref{eq:langevin-dimless} by noting that
directed transport is possible in certain parameter ranges even in the
deterministic case where noise is absent. However, to probe the
robustness of transport, we consider in the following primarily the limit of
weak noise by choosing $\hat T=10^{-3}$. In fact, it has been shown
for underdamped ratchets with time-independent potentials and periodic
driving that even an infinitesimal amount of noise can change the
rectification from chaotic to stable \cite{Marchesoni:2006a}.  To look
for similar generic behavior of our pulsating ratchet, we consider a
specific geometry consisting of a symmetric and a sawtooth-like
surface profile corresponding to three harmonics with $f_1=0.0492$,
$f_2=0.0241$, $f_3=0.0059$ and $f_n=0$ for $n>3$. Inset (a) of
Fig.~\ref{Fig:force} shows these profiles in their stable position
with $\hat b=0.182$ that minimizes the Casimir energy. The resulting
spatial variation of the Casimir force with $\hat b$ is plotted in
Fig.~\ref{Fig:force} for minimal ($s=0$) and maximal ($s=\pi/\omega$)
surface distance with parameters $H_0/\lambda=0.1$, $\eta=0.65$. It
can be clearly seen that the asymmetry is reduced at larger distance
where the variation of the force becomes more sinusoidal. Inset (b)
shows the on-off-like time-dependence of the force amplitude at $\hat
b=0$ due to the oscillating surface distance.

The non-linear equation of motion of Eq.~\eqref{eq:langevin-dimless}
has to be solved numerically. The trajectory $\hat b(s)$ was obtained
from a second order Runge-Kutta algorithm. As initial conditions we
used an equidistant distribution over the interval $[-1,1]$ for $\hat
b(0)$ and $\dot{\hat b}(0)=0$. For each set of parameters we
calculated 200 different trajectories from varying initial conditions
and noise, each evolving over $4 \times 10^3$ periods $2\pi/\omega$ so
that transients have decayed. The average velocity $\langle\!\langle v
\rangle\!\rangle$ involves two different averages of $\dot{\hat
  b}(s)$: The first average is over initial conditions and noise for
every time step, then the averaged trajectory is averaged over all
discrete times of the numerical solution. For an efficient directed
transport it is not sufficient to have only a finite average
$\langle\!\langle v \rangle\!\rangle$. To exclude trajectories with a
high number of velocity reversals, the fluctuations about the average
velocity must be small, i.e., the variance $\sigma^2 =
\langle\!\langle v^2 \rangle\!\rangle -\langle\!\langle v
\rangle\!\rangle^2$ must be smaller than
$\langle\!\langle v \rangle\!\rangle^2$.


\begin{figure}
\includegraphics[width=0.85\linewidth]{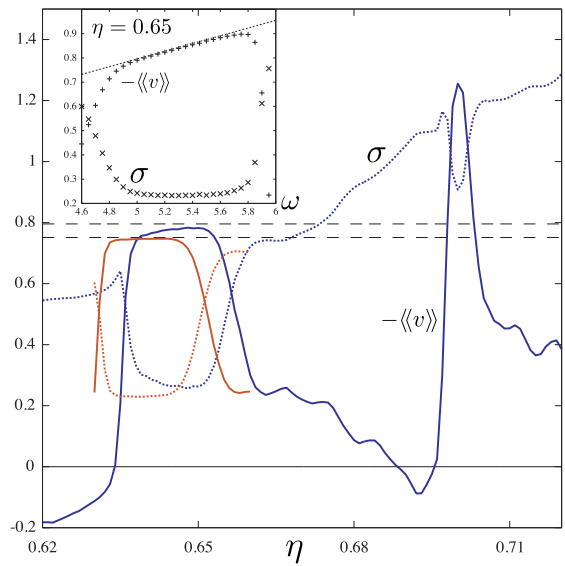}
\vspace*{-0.2cm}
\caption{\label{Fig:v-of-eta} Mean $\langle\!\langle v
  \rangle\!\rangle$ and standard deviation $\sigma$ of the (negative)
  velocity as function of the driving amplitude $\eta$ for the
  frequencies $\omega=5.0$ and $\omega=4.72$ (for the latter only the
  stable plateau is shown). The parameters are $H_0=0.1\lambda$,
  $\hat\gamma=0.9$, $\hat T=10^{-3}$. Inset: Dependence of the same
  quantities on frequency for fixed $\eta=0.65$. Straight dashed lines
  correspond in both graphs to the velocity $\omega/(2\pi)$.}
\vspace*{-0.5cm}
\end{figure}

Naively, one can expect directed motion of the surface profile
$h_2(x_1)$ into the positive $x_1$-direction ($\dot{\hat b}<0$) since
the Casimir force in Fig.~\ref{Fig:force} is asymmetric with negative
values lasting for longer time than positive ones. However, the actual
behavior is more complicated due to chaotic
dynamics. Fig.~\ref{Fig:v-of-eta} shows the dependence of the average
velocity and its standard deviation $\sigma$ on the driving amplitude
$\eta$ and frequency $\omega$ for $H_0=0.1 \lambda$, $\hat\gamma=0.9$.
For a fixed frequency there is an optimal interval of driving
amplitudes across which the average velocity is almost constant with
$\langle\!\langle v \rangle\!\rangle \simeq -\omega/(2\pi)$. Small
deviations from the latter value result from noise as we have checked
by studying the dynamics at $\hat T=0$. At higher driving amplitudes
we observe a second narrower interval with maximal $\langle\!\langle v
\rangle\!\rangle$ which is more strongly reduced and smeared out from
its deterministic value $-2\times \omega/(2\pi)$ by noise. At the
plateaus of constant velocity the standard deviation $\sigma$ is
substantially reduced, rendering transport efficient. Outside the
plateaus velocity reversals occur and $\sigma$ increases linearly with
$\eta$. For fixed amplitude $\eta$, the average velocity is stable at
the value $-\omega/(2\pi)$ over a sizeable frequency range (see inset
of Fig.~\ref{Fig:v-of-eta}).

In order to understand the observed behavior we have analyzed
the dynamics in the three dimensional extended phase space. There
attractors of the long-time dynamics can be identified from Poincar\'e
sections using the period $2\pi/\omega$ of the surface oscillation as
stroboscopic time. To obtain a compact section, the trajectory is
folded periodically in $x_1$ on one period of the Casimir potential.
From these sections we can distinguish between periodic and chaotic
orbits. As a start, we consider the deterministic limit with $\hat
T=0$. The plateaus around $\eta=0.65$ and $\eta=0.7$ result both from
periodic orbits of period one, corresponding to a single point in the
Poincar\'e section. On the right (downward) edges of the first
plateaus we have observed period doubling, i.e., a periodic attractor
with period two. Upon a further increase of $\eta$, chaotic orbits
dominate the motion. Hence the system exhibits a period-doubling route
to chaos with enhanced velocity fluctuations. The findings apply
basically also to weak noise ($\hat T=10^{-3}$) but the sharp points
of the periodic attractors in the Poincar\'e sections are smeared out
leading to a decreased $\langle\!\langle v \rangle\!\rangle$.  The
transition from chaotic to periodic dynamics at the beginning of the
rising edge of the plateaus is accompanied by a velocity reversal.
This is consistent with the earlier observation for non-pulsating
potentials that velocity reversals are due to a bifurcation from
chaotic to periodic dynamics \cite{Mateos:2000a}.

\begin{figure}
\includegraphics[width=1.\linewidth]{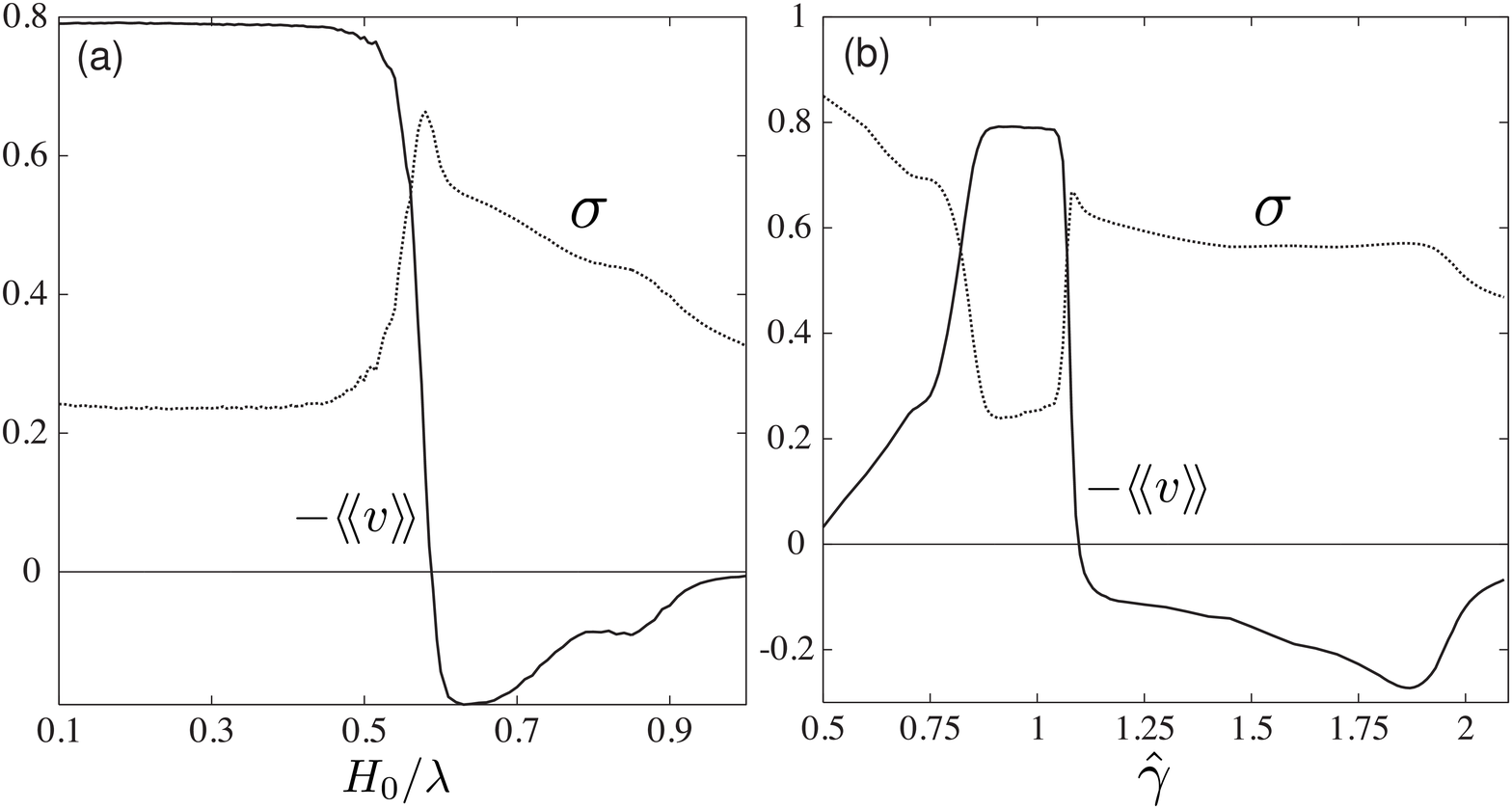}
\vspace*{-0.6cm}
\caption{\label{Fig:v-of-H+gamma} Mean $\langle\!\langle v
  \rangle\!\rangle$ and standard deviation $\sigma$ of the (negative)
  velocity as function of (a) the mean plate distance $H_0$ for
  $\hat\gamma=0.9$ and (b) damping $\hat\gamma$ for
  $H_0=0.1\lambda$. The other parameters are $\eta=0.65$,
  $\omega=5.0$, $\hat T=10^{-3}$.}
\vspace*{-0.5cm}
\end{figure}

The amplitude of the Casimir potential can be tuned by varying the
mean distance $H_0$. From Fig.~\ref{Fig:v-of-H+gamma}(a) we see that
the dynamics show a sharp transition at a critical $H_0/\lambda$ from
efficient transport with large $\langle\!\langle v \rangle\!\rangle$
and small $\sigma$ to chaotic dynamics with vanishing velocity. The
transition is accompanied by a velocity reversal and peaked velocity
fluctuations. Interestingly, below the transition $\langle\!\langle v
\rangle\!\rangle$ is almost constant independently of $H_0/\lambda$.
The observed transport behavior is also stable against a change of
effective damping $\hat\gamma$ as shown in
Fig.~\ref{Fig:v-of-H+gamma}(b). Whereas fluctuations increase with
decreasing $\hat\gamma$, there is a stable plateau of constant average
velocity across which fluctuations are diminished. In the
deterministic limit, we have also observed additional plateaus with
inverted and doubled average velocity by varying $\hat\gamma$ and
$\eta$. Remnants of a second plateau around $\hat\gamma=1.9$, washed
out by noise, can be seen in Fig.~\ref{Fig:v-of-H+gamma}(b).


Finally, let us estimate typical velocities $v_0=\lambda/\tau$.  With
the typical lengths $H_0=0.1\mu$m, $a=10$nm realized in recent Casimir
force measurements \cite{Chen:2002a} and an area mass density of
$\rho=10$g/m$^2$ for silicon plates with a thickness of a few microns,
one obtains $v_0=\sqrt{\hbar c a^2/\rho H_0^5}\approx 5.5$mm/s. The
actual average velocity $v_0 \omega/2\pi$ is of the same order for the
frequencies studied above. For $\lambda=1\mu$m, the time scale is
$\tau=\lambda/v_0\approx 10^{-4}$s leading to driving frequencies and
damping rates in the kHz range for the parameters considered here.

Our results show that Casimir interactions offer novel contact-less
translational actuation schemes for nanomechanical systems. Similar
ratchet-like effects are expected between objects of different shapes
as, e.g., periodically structured cylinders, inducing rotational
motion. The use of fluctuation forces appear also promising to move
nano-sized objects immersed in a liquid where electrostatic actuation
is not possible. Another application is the separation and detection
of particles of differing mass adsorbed to the surfaces.  For surfaces
oscillating at very high frequencies additional interesting phenomena
related to the dynamical Casimir effect occur
\cite{Golestanian:1997a}, leading to the emission of photons that
could contribute to ratchet-like effects as well.

I acknowledge useful discussions with R. Golestanian and M. Kardar.

\end{document}